\documentclass[12pt]{iopart}
\usepackage{graphicx}
\begin{document}
\title{Derivatives and inequalities for order parameters in the Ising spin glass}
\author{Hidetoshi Nishimori}
\address{Department of Physics, Tokyo Institute of Technology,
Oh-okayama, Meguro-ku, Tokyo 152-8551, Japan}
\begin{abstract}
Identities and inequalities are proved for the
order parameters, correlation functions and their derivatives of the Ising spin glass.
The results serve as additional evidence that the ferromagnetic phase is
composed of two regions, one with strong ferromagnetic ordering and
the other with the effects of disorder dominant.
The Nishimori line marks a crossover between these two regions.
\end{abstract}
%

\maketitle

\section{Introduction}
One of the outstanding problems in the theory of finite-dimensional spin
glasses is the structure of the phase diagram.
Numerical investigations have revealed the precise locations of
critical points and phase boundaries \cite{numerical}.
However we still have only limited knowledge from analytical treatments
of the problem.
An interesting exception is the gauge theory which makes use of gauge
symmetry of the system to derive a variety of exact/rigorous results
on physical quantities including the energy, specific heat and
correlation functions \cite{Nishimori81,Nishimori-b}.
In the present paper we derive a class of relations for the order parameters
and correlation functions using the gauge theory to clarify the
behaviour of the system within the ferromagnetic phase.

Properties of the ferromagnetic phase in models of spin glasses have not
been studied very extensively compared to the spin glass phase.
Nevertheless, as shown in references \cite{Nishimori-Sollich}
and \cite{Nishimori93},
a very interesting non-trivial change of the system behaviour is
observed on a line, the Nishimori line, in the phase diagram:
The spins become more ferromagnetically ordered ({\it i.e.}
parallel to each other)
as the temperature is lowered above this line whereas the same
spins turn to become
misaligned below the same line with further decrease of the temperature.
Although this line is not a phase boundary
in the thermodynamic sense, it marks in the above sense
a clear crossover between the two regions within the ferromagnetic phase.
The argument in the present paper using the gauge theory reinforces
this picture through relations between the order parameters and their
derivatives.

An important consequence of the gauge theory is an identity between the
distribution function of the ferromagnetic order parameter $P_m(x)$
and that of the spin glass order parameter $P_q(x)$: These two functions
are equal to each other $P_m(x)=P_q(x)$ on the Nishimori line
\cite{Nishimori-b,NS}.
This functional identity implies the absence of replica symmetry breaking
because $P_m(x)$ is trivial, composed of at most two simple delta functions,
and, therefore, so is $P_q(x)$.
The relation $P_m(x)=P_q(x)$ also leads to
the equality $m=q$, an identity
between the ferromagnetic and spin glass order parameters
\cite{Nishimori81,Nishimori-b}, implying an {\it exact} balance between the two
types of ordering.
From $m=q$ we may expect that the ferromagnetic order parameter exceeds
the spin glass order parameter $m>q$ above the Nishimori line
because, in the limit of non-random system (which is above the line),
we have $q=m^2<m$.
Another reason to expect $m>q$ is that, as mentioned above,
ferromagnetic ordering dominates above the line.
The opposite inequality $q>m$ is likely to hold below the same line
since the spin glass phase (lying below the line) has $q>m=0$,
and, in addition, the effects of quenched randomness is more dominant
(apparent misalignment of spins)
below the line as explained above.

These two inequalities for the order parameters can be verified
in the mean-field Sherrington-Kirkpatrick model \cite{SK}
within the replica-symmetric solution (which is valid
near the Nishimori line as mentioned above) but have been considered
difficult to check analytically
for general finite-dimensional systems.
We show in the present paper that these inequalities are closely
related with temperature derivatives and some inequalities of the order parameters,
thus coming closer to a proof.

We present our results and their proofs in the next section.
Discussions are given in the last section.
Some of the details of calculations are described in the Appendix.

\section{Identities and inequalities}

\begin{figure}
  \begin{center}
  \includegraphics[width=.28\linewidth]{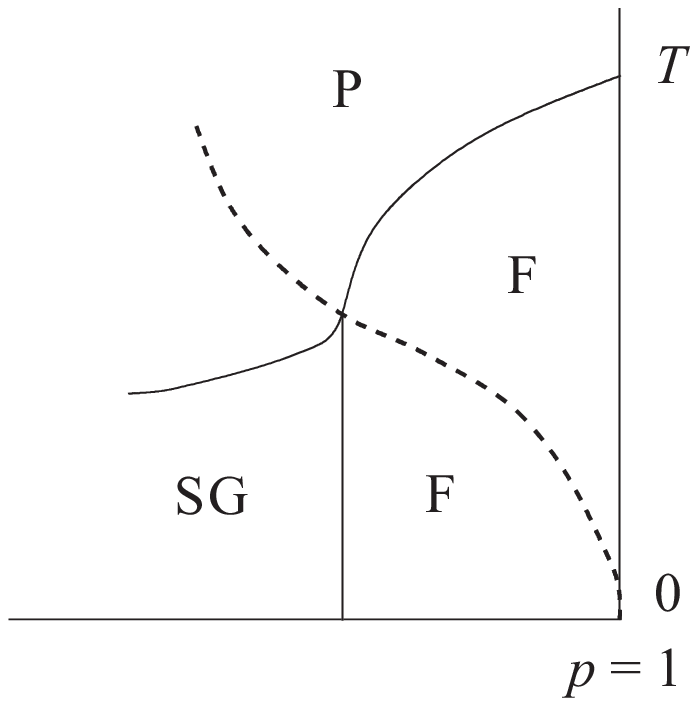}
  \end{center}
 \caption{A plausible generic phase diagram of the $\pm J$ model. The system has
  paramagnetic (P), ferromagnetic (F) and spin glass (SG) phases.
  The Nishimori line $K=K_p$ shown dashed marks a crossover
   between two regions within the ferromagnetic phase.}
 \label{fig:phase-diagram}
\end{figure}

To be specific, let us consider the $\pm J$ Ising model on an arbitrary lattice
with the probability of ferromagnetic interaction denoted by $p$.
The expected phase diagram
is depicted in figure \ref{fig:phase-diagram}.
The main physical quantities we treat in this paper are the
ferromagnetic and spin glass order parameters defined by
\begin{equation}
   m=[ \langle S_i \rangle_K ], ~~~
   q=[\langle S_i \rangle_K^2 ], 
   \label{mq-def}
\end{equation}
respectively, with $i$ well interior of the lattice.
The inner brackets denote the thermal average with coupling constant
$K=J/T$, and the outer square brackets are for the configurational
average characterized by the parameter $p$.
The spins on the boundary of the lattice under consideration are
set to up states to avoid trivial vanishing of the thermal expectation
value of the one-point function $\langle S_i \rangle_K$.
It is sufficient to consider a single parameter $q$ as a
spin glass order parameter instead of its distribution function $P_q(x)$
because replica symmetry breaking is absent when $K=K_p\equiv
\frac{1}{2}\log ( p/(1-p))$ \cite{NS,Nishimori-b}, the Nishimori line,
on and near which we concentrate ourselves for the moment.

The identity $m=q$ valid when $K=K_p$ has long been known \cite{Nishimori81}.
The first of our new results is the following relations:
 \begin{eqnarray}
   \frac{\partial q}{\partial K}=2 \frac{\partial m}{\partial K}
         \label{1st-deriv}\\
    \frac{\partial^2 q}{\partial K^2}\ge 2 \frac{\partial^2 m}{\partial K^2},
         \label{2nd-deriv}
 \end{eqnarray}
both of which hold under the condition $K=K_p$.

The proof is straightforward.
As shown in the Appendix, the magnetization is rewritten using gauge
transformation as
\begin{equation}
   m=[\langle S_i \rangle_K]
  =[\langle \sigma_i \rangle_{K_p} \langle S_i \rangle_K],
   \label{magnetization}
\end{equation}
where $\langle \sigma_i \rangle_{K_p}$ is the thermal average of the Ising
spin $\sigma_i$ (introduced by the gauge transformation) of the same system as
the original $\pm J$ model with effective coupling $K_p$.
The first and second derivatives of $m$ are then
\begin{eqnarray}
    \frac{\partial m}{\partial K}=\left[
      \langle \sigma_i \rangle_{K_p} \frac{\partial}{\partial K}
                \langle S_i \rangle_K \right],
       \label{m-1st}\\
    \frac{\partial^2 m}{\partial K^2}=\left[
      \langle \sigma_i \rangle_{K_p} \frac{\partial^2}{\partial K^2}
                \langle S_i \rangle_K \right].
       \label{m-2nd}
\end{eqnarray}
The derivatives of the spin glass order parameter are obtained directly from
the definition (\ref{mq-def}):
  \begin{eqnarray}
    \frac{\partial q}{\partial K}=2 \left[
      \langle S_i \rangle_{K} \frac{\partial}{\partial K}
                \langle S_i \rangle_K \right],
       \label{q-1st}\\
    \frac{\partial^2 q}{\partial K^2}=2\left[
      \langle S_i \rangle_{K} \frac{\partial^2}{\partial K^2}
                \langle S_i \rangle_K \right]
       +2\left[\left( \frac{\partial}{\partial K}  \langle S_i \rangle_K 
  \right)^2 \right].
       \label{q-2nd}
\end{eqnarray}
The identity $m=q$ for $K=K_p$ immediately follows from (\ref{magnetization})
and (\ref{mq-def}) because $\langle \sigma_i \rangle_{K_p}=
\langle S_i \rangle_K$ if $K=K_p$.
The identity (\ref{1st-deriv}) valid for $K=K_p$ is also easy to verify
from (\ref{m-1st}) and (\ref{q-1st}).
The inequality (\ref{2nd-deriv}) is a consequence of (\ref{m-2nd})
and (\ref{q-2nd}).

Similar relations hold for more general correlation functions.
Let us define two correlations:
  \begin{equation}
   C_{ijk\cdots}^{(2l+1)}=[\langle S_i S_j S_k \cdots \rangle_K^{2l+1} ],~~~
  C_{ijk\cdots}^{(2l+2)}=[\langle S_i S_j S_k \cdots \rangle_K^{2l+2} ],
\end{equation}
where $l$ is a non-negative integer and $\{i, j, k,\cdots\}$ is an arbitrary
set of sites.

This $C_{ijk\cdots}^{(2l+1)}$ satisfies the following identity (see the Appendix)
\begin{equation}
C_{ijk\cdots}^{(2l+1)}=[\langle \sigma_i \sigma_j \sigma_k \cdots \rangle_{K_p} 
\langle S_i S_j S_k \cdots \rangle_K^{2l+1} ].
  \label{C-identity}
\end{equation}
Using the fact that $C_{ijk\cdots}^{(2l+2)}$ is gauge invariant,
we can prove the following relations at $K=K_p$:
\begin{equation}
     C_{ijk\cdots}^{(2l+1)}=C_{ijk\cdots}^{(2l+2)},~~~
   \frac{1}{2l+1}\frac{\partial}{\partial K}C_{ijk\cdots}^{(2l+1)}
     =\frac{1}{2l+2}\frac{\partial}{\partial K}C_{ijk\cdots}^{(2l+2)}.
     \label{CC_relation}
\end{equation}
No simple relation exists between the second derivatives for general $l$.
It is to be noted that equation (\ref{CC_relation}) holds not just in the
ferromagnetic phase but in the paramagnetic phase as well
whereas equations (\ref{1st-deriv}) and (\ref{2nd-deriv}) are trivial
in the paramagnetic phase as both sides vanish identically.

We next discuss our second new result for the order parameters.
Let us denote the dependence of the order parameters on the temperature and
probability parameter explicitly as
$m(K, K_p)$ and $q(K, K_p)$.
Then it is possible to show that
 \begin{eqnarray}
   &&m(K, K_p)\ge q(K, K_p) 
   \nonumber\\
   &&\Rightarrow m(K_p, K_p)\ge m(K, K_p) ~~
    {\rm and}~~q(K_p, K_p)\ge q(K, K_p)
    \label{mq_ineq2}
 \end{eqnarray}
for any $K$ and $K_p$.
To prove this, it is useful to write (\ref{magnetization}) explicitly as
 \begin{equation}
   m(K, K_p)=\sum_k P(k) \langle \sigma_i \rangle_{K_p}^{(k)}
    \langle S_i \rangle_{K}^{(k)}
     =\sum_k \sqrt{P(k)} \langle \sigma_i \rangle_{K_p}^{(k)}\cdot \sqrt{P(k)} 
    \langle S_i \rangle_{K}^{(k)},
     \label{m_expl}
 \end{equation}
where $k$ stands for a bond configuration.
Let us square both sides of the above equation and apply the Schwarz inequality to obtain
 \begin{eqnarray}
   m(K, K_p)^2 &\le &\sum_k P(k) (\langle \sigma_i \rangle_{K_p}^{(k)})^2
    \sum_k P(k) (\langle S_i \rangle_{K}^{(k)})^2
     \nonumber\\
      &=&[\langle \sigma_i \rangle_{K_p}^2][\langle S_i \rangle_{K}^2]
     \nonumber\\
       &=& q(K_p, K_p) q(K, K_p).
      \label{ineq2}
 \end{eqnarray}
Now, if we assume $m(K, K_p) \ge q(K, K_p)$, then $q(K, K_p)$ on the right hand
side can be replaced by $m(K, K_p)$ to yield
  \begin{equation}
    m(K, K_p)^2 \le q(K_p, K_p)m(K, K_p).
      \label{ineq3}
   \end{equation}
Since we are considering the ferromagnetic phase with up-spin boundaries, we
have $m(K, K_p)>0$, and therefore by dividing both sides by $m(K, K_p)$, we find
 \begin{equation}
    m(K, K_p) \le q(K_p, K_p) =m(K_p, K_p).
  \end{equation}
This is the first half of the result (\ref{mq_ineq2}).

The second half is proved similarly. From (\ref{ineq2}) and the assumption
$m(K, K_p) \ge q(K, K_p)$, we find
  \begin{equation}
    q(K, K_p)^2 \le q(K_p, K_p)q(K, K_p)
  \end{equation}
and thus
  \begin{equation}
    q(K, K_p) \le q(K_p, K_p).
  \end{equation}

A generalization to correlation functions is straightforward.
The result is
 \begin{eqnarray}
   && C_{ijk\cdots}^{(2l+1)}(K, K_p)\ge C_{ijk\cdots}^{(4l+2)}(K, K_p) 
    \nonumber\\
   &&\Rightarrow 
   C_{ijk\cdots}^{(1)}(K_p, K_p)\ge C_{ijk\cdots}^{(2l+1)}(K, K_p) 
    ~{\rm and}~
    C_{ijk\cdots}^{(2)}(K_p, K_p)\ge C_{ijk\cdots}^{(4l+2)}(K, K_p) .
    \nonumber\\
   && \label{mq_ineq2a}
 \end{eqnarray}
To prove these inequalities, we apply the gauge transformation and Schwarz
inequality to $C_{ijk\cdots}^{(2l+1)}(K, K_p)$:
  \begin{eqnarray}
    C_{ijk\cdots}^{(2l+1)}(K, K_p)^2 &=&[\langle S_i S_j S_k\cdots \rangle_K^{2l+1}]^2
      \nonumber\\
 &=&\left( \sum_k \sqrt{P(k)} \langle \sigma_i \sigma_j \sigma_k\cdots \rangle_{K_p}^{(k)}
  \cdot \sqrt{P(k)} 
  \left(\langle S_i S_j S_k\cdots \rangle_{K}^{(k)}\right)^{2l+1}
    \right)^2
    \nonumber\\
  &\le &
 \sum_k P(k) \left( \langle \sigma_i \sigma_j \sigma_k\cdots \rangle_{K_p}^{(k)}\right)^2
 \sum_k P(k) \left( \langle S_i S_j S_k\cdots \rangle_{K}^{(k)}\right)^{4l+2}
  \nonumber\\
  &=&
   [\langle \sigma_i \sigma_j \sigma_k\cdots \rangle_{K_p}^2]
   [\langle S_i S_j S_k\cdots \rangle_{K}^{4l+2}]
   \nonumber\\
   &=&
    C_{ijk\cdots}^{(2)}(K_p, K_p) C_{ijk\cdots}^{(4l+2)}(K, K_p).
    \label{corr1}
  \end{eqnarray}
When $C_{ijk\cdots}^{(2l+1)}(K, K_p)\ge C_{ijk\cdots}^{(4l+2)}(K, K_p)$, the second
factor on the right hand side of (\ref{corr1}) is bounded from above by
$C_{ijk\cdots}^{(2l+1)}(K, K_p)$ to yield
  \begin{equation}
    C_{ijk\cdots}^{(2l+1)}(K, K_p)^2 \le C_{ijk\cdots}^{(2)}(K_p, K_p)
     C_{ijk\cdots}^{(2l+1)}(K, K_p).
  \end{equation}
Since $C_{ijk\cdots}^{(2l+1)}(K, K_p)> 0$ under the up-spin boundary condition,
we have
  \begin{equation}
    C_{ijk\cdots}^{(2l+1)}(K, K_p) \le C_{ijk\cdots}^{(2)}(K_p, K_p)
      =C_{ijk\cdots}^{(1)}(K_p, K_p),
   \end{equation}
the final inequality being a result of gauge transformation of the
kind described in the Appendix.
This ends the proof of the first inequality of (\ref{mq_ineq2a}).
By replacing the left hand side of (\ref{corr1}) with the lower bound
$C_{ijk\cdots}^{(4l+2)}(K, K_p)^2$, we arrive at the second relation
 \begin{equation}
    C_{ijk\cdots}^{(4l+2)}(K, K_p) \le  C_{ijk\cdots}^{(2)}(K_p, K_p).
 \end{equation}

It is possible to apply similar arguments to the
other models of spin glasses with gauge symmetry including
the random Ising model with Gaussian-distributed interactions
and $XY$ gauge glass \cite{Nishimori-b,Ozeki-Nishimori}.
The physical significance of the results obtained in this section
will be discussed in the next section.
%
\section{Discussions}
\begin{figure}
  \begin{center}
  \includegraphics[width=.55\linewidth]{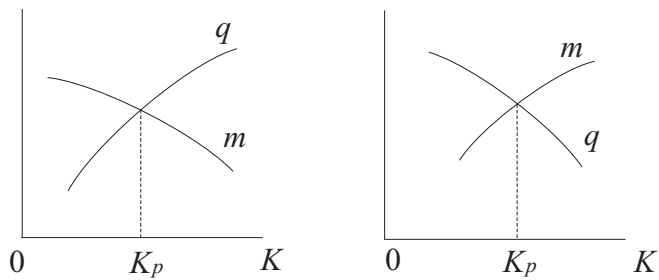}
  \end{center}
  \caption{It is forbidden that the derivatives of $m$ and $q$ have
  different signs at $K=K_p$ as shown in this figure.}
 \label{fig:excluded}
\end{figure}

\begin{figure}
  \begin{center}
  \includegraphics[width=.55\linewidth]{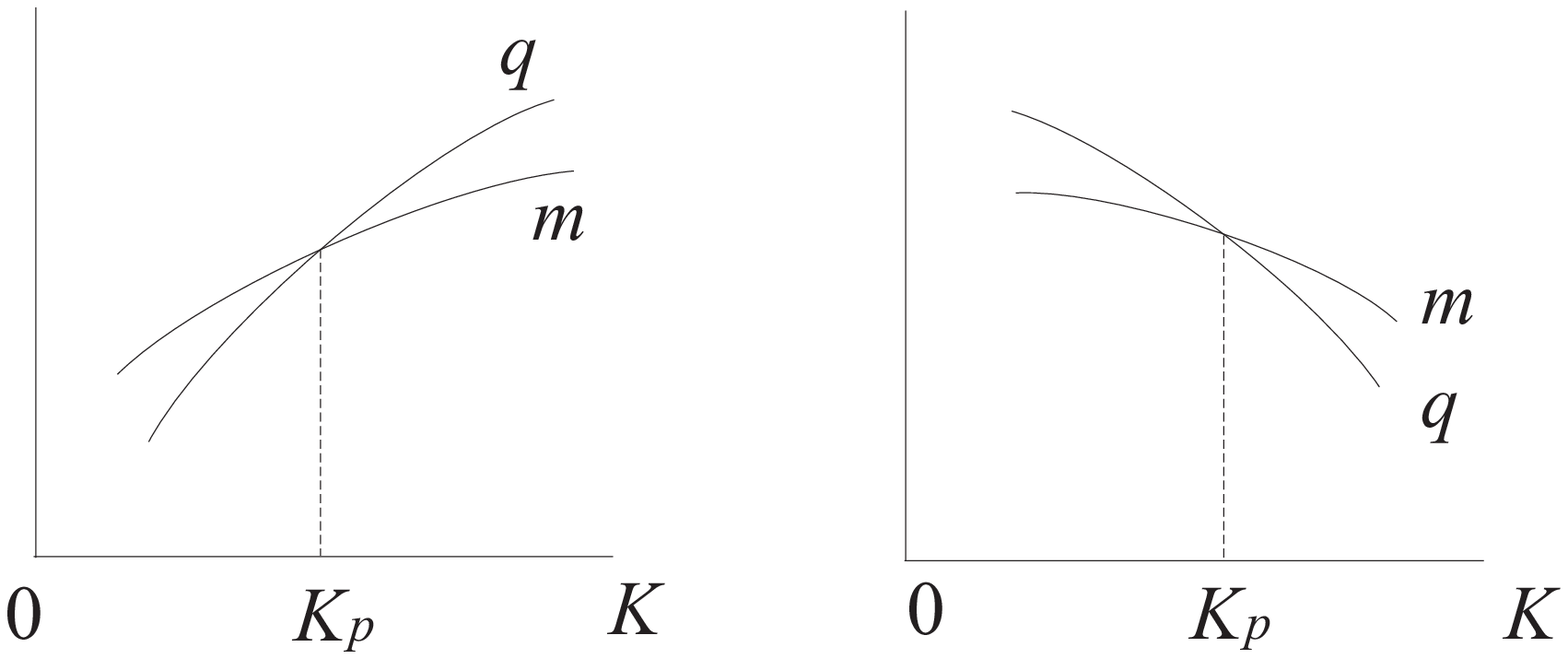}
  \end{center}
  \caption{The spin glass order parameter increases twice as rapidly as
   the ferromagnetic order parameter around $K=K_p$
   if these quantities are increasing functions of $K$ (left).
   The opposite possibility is shown on the right.}
 \label{fig:case1}
\end{figure}

An immediate consequence of (\ref{1st-deriv}) is that the derivatives
of $q$ and $m$ have the same sign when $K=K_p$.
It is forbidden that, for example, the spin glass order parameter
$q$ increases whereas the ferromagnetic order parameter $m$ decreases
as depicted in figure \ref{fig:excluded}.
%
In the plausible case that $\partial q/\partial K >0$, it follows from
(\ref{1st-deriv}) that $q$ increases twice as rapidly as $m$ does
as the temperature is lowered (see figure \ref{fig:case1} (left)).
It naturally follows that $q$ is larger than $m$ at least slightly
below the Nishimori line $(K>K_p)$ and the opposite inequality
$m>q$ holds when $K<K_p$.
This is the most natural case as discussed in the introduction:
The ferromagnetic ordering is dominant $(m>q)$ above the Nishimori line
($K<K_p$)
and quenched-disorder-driven random ordering proliferates in the sense $q>m$
below the same line ($K>K_p$).
The point is that we have reduced the inequality $q>m$
for $K>K_p$ to an intuitively natural relation $\partial q/\partial K>0$
at $K=K_p$ (although we still do not have a rigorous proof of the latter.)

If the opposite inequality $\partial q/\partial K<0$
holds at $K=K_p$, the order parameters behave as depicted in
figure \ref{fig:case1} (right).
We cannot exclude this possibility from the present
argument, but it seems quite unlikely
that both order parameters $q$ and $m$ decrease with temperature
decrease on and around the line $K=K_p$ which runs through the central part
of the ferromagnetic phase as seen in figure \ref{fig:phase-diagram}.
If there exists a reentrant transition (below which $m$ vanishes
but $q$ can stay finite), only $m$ may decrease toward such a
transition temperature from above unlike figure \ref{fig:case1} (right).
Although we believe that such a situation is not plausible to exist
\cite{Nishimori-geometry}, it would happen at very low
temperatures if it does at all, not around the line $K=K_p$
which runs through relatively high temperature parts of the
ferromagnetic phase.

The final possibility is that the derivatives in (\ref{1st-deriv}) vanish.
Again, both derivatives should vanish, not just one of them.
The inequality for the second derivative (\ref{2nd-deriv}) does not
tell much about the behaviour of $m$ and $q$ around $K=K_p$.

It is instructive to remember in this connection
that the average {\it sign} of local
magnetization reaches its maximum at $K=K_p$ as a function of $K$
(or the temperature) \cite{Nishimori93}
 \begin{equation}
    M(K, K_p)\equiv [{\rm sgn}\langle S_i \rangle_K] \le M(K_p, K_p).
   \label{align}
 \end{equation}
This means that the number of up spins under up-spin boundaries becomes
maximum at $K=K_p$ as a function of $K$.
Although this relation (\ref{align}) can be proved for an arbitrary $K$
by the gauge theory \cite{Nishimori93}, it is useful to check it
by taking the derivative of (\ref{align}) (as we did for $m$ and $q$):
 \begin{equation}
  \fl
       \frac{\partial M}{\partial K}=
  \left[ \frac{\partial}{\partial K}{\rm sgn}\langle S_i \rangle_K\right]
 =2 \left[ \delta (\langle S_i \rangle_K)
    \frac{\partial}{\partial K}\langle S_i \rangle_K\right]  
  =2 \left[\langle \sigma_i \rangle_{K_p}   \delta (\langle S_i \rangle_K)
    \frac{\partial}{\partial K}\langle S_i \rangle_K\right]  =0,
   \label{M-deriv}
 \end{equation}
the last equality being valid for $K=K_p$.
Equation (\ref{M-deriv}) implies that the effects of spins with
positive temperature derivative ($\partial \langle S_i \rangle_K/
\partial K >0$) just match those of
negative temperature derivative ($\partial \langle S_i \rangle_K/
\partial K <0$) at $K=K_p$ if we concentrate ourselves on the
spins with vanishing local magnetization $\delta (\langle S_i \rangle_K)$.
Thus, at $K=K_p$, some spins change its local magnetization from positive value
to negative value whereas essentially the same number of spins change
the sign of local magnetization in the opposite way.
It should be noted, however, that this observation does not necessarily mean
vanishing derivatives of the ferromagnetic and spin glass order parameters
at $K=K_p$, that is, $\partial m/\partial K$ and $\partial q/\partial K$
are in general not vanishing at $K=K_p$:
The absolute value of local magnetization $|\langle S_i \rangle_K|$,
which is ignored in $M(K)$, at sites with $\langle S_i \rangle_K>0$
may grow more rapidly than at sites with $\langle S_i \rangle_K<0$
as $K$ decreases, compensating for the decrease in the number of
up spins below $K=K_p$, leading to a positive derivative
$\partial m /\partial K>0$.

The present argument does not apply in the paramagnetic
phase where  the order parameters $m$ and $q$ vanish.
However, the relation (\ref{CC_relation}) for correlation functions, the two-point
functions $C_{ij}^{(2l+1)}(K, K_p)$ and $C_{ij}^{(2l+2)}(K, K_p)$ in particular,
suggests that $C_{ij}^{(2l+1)}(K, K_p)>C_{ij}^{(2l+2)}(K, K_p)$ if $K<K_p$ and
$C_{ij}^{(2l+1)}(K, K_p)<C_{ij}^{(2l+2)}(K, K_p)$ if $K>K_p$.
This means that the ferromagnetic
correlation length $\xi_m$ (defined by $C_{0r}^{(1)}\approx \exp (-r/\xi_m), r\gg 1$)
is larger than the spin glass correlation length $\xi_q$
(defined by $C_{0r}^{(2)}\approx \exp (-r/\xi_q), r\gg 1$)
for $K<K_p$ (above the Nishimori line) whereas the opposite inequality holds below
the line.

Very similar conclusions follow from the result (\ref{mq_ineq2}).
The relation between $m$ and $q$ shown in figure \ref{fig:excluded} violates
these inequalities because, in the temperature range where $m(K, K_p)$
exceeds $q(K, K_p)$, $m(K, K_p)$ is seen to be larger than its
value at $K=K_p$.
The cases given in figure \ref{fig:case1} are compatible with the
present inequalities:
If $m(K, K_p)> q(K, K_p)$, then $m(K, K_p)$ is smaller than $m(K_p, K_p)$.
An advantage of the present inequalities (\ref{mq_ineq2}) over the
relation (\ref{1st-deriv})
is that we can analyze the behaviour of order parameters well away
from the Nishimori line, that is, $K$ can be much larger or smaller than $K_p$.
A weak point is that quantitative relations for the increase/decrease rates
are not given unlike (\ref{1st-deriv}).
Another problem to remember concerning (\ref{mq_ineq2})
is that we may not be able to describe the
system only in terms of $m$ and $q$ at very low temperatures
if replica symmetry breaking exists as it is the case in the
Sherrington-Kirkpartrick model \cite{AT}.

The analysis presented above strongly indicates that the Nishimori line $K=K_p$
marks a crossover between the purely ferromagnetically-ordered region
and disorder-dominated region within the ferromagnetic phase.
This observation is consistent also with renormalization group analyses:
The low-temperature region is controlled by a fixed point at $T=0$
with finite disorder ($p<1$) whereas the high-temperature region is described
by a different fixed point representative of the critical curve
above the multicritical point \cite{Doussal,Hukushima}.

An important future direction of investigation is a rigorous proof
of the relation $\partial m/\partial K>0$ at $K=K_p$, which
needs additional ideas.
\section*{Appendix}
In this Appendix we drive equations (\ref{magnetization}) and
(\ref{C-identity}) following references \cite{Nishimori81}
and \cite{Nishimori-b}.
The ferromagnetic order parameter is defined by
 \begin{equation}
 \fl
  m=[\langle S_i \rangle_K ]  
  = \frac{1}{(2\cosh K_p)^{N_B}}
  \sum_{\{\tau_{ij}=\pm 1\}} \exp (K_p \sum_{\langle ij\rangle} \tau_{ij})
   \frac{\sum_{\{S\}} S_i \exp (K\sum_{\langle ij\rangle} \tau_{ij}S_i S_j)}
  {\sum_{\{S\}} \exp (K\sum_{\langle ij\rangle} \tau_{ij}S_i S_j)}.
      \label{m-def-detail}
  \end{equation}
Here $\tau_{ij}$ is the sign of $J_{ij}$ ($\tau_{ij}=J_{ij}/|J_{ij}|=\pm 1$),
the sums in the exponents run over all pairs of sites on the lattice under
consideration, and $N_B$ is the total number of bonds.
The factor $ \exp (K_p  \tau_{ij})/2\cosh K_p$ gives the probability
weight of configurational average of the $\pm J$ model
since this quantity equals
$p$ for $\tau_{ij}=1$ and $1-p$ for $\tau_{ij}=-1$ as can be verified
from the definition $K_p=\frac{1}{2}\log (p/(1-p))$.
Spins on the boundary are all up.

Let us apply a gauge transformation
 \begin{equation}
      S_i\to S_i \sigma_i,~~~\tau_{ij}\to\tau_{ij}\sigma_i \sigma_j
    \label{g-trsf}
 \end{equation}
to all sites,
where $\sigma_i$ is a gauge variable fixed either to $1$ or $-1$
arbitrarily at each site ($+1$ on the boundary).
The Hamiltonian in the exponents of (\ref{m-def-detail}) is invariant under
this gauge transformation.
Since the gauge transformation is just a re-definition of running variables
in (\ref{m-def-detail}), it does not affect the value of the right hand side,
and we have
 \begin{equation} 
 \fl
  m= \frac{1}{(2\cosh K_p)^{N_B}}
  \sum_{\{\tau_{ij}\}} \exp (K_p \sum_{\langle ij\rangle} \tau_{ij}
   \sigma_i \sigma_j) 
   \frac{\sigma_i\sum_{\{S\}} S_i
      \exp (K\sum_{\langle ij\rangle} \tau_{ij}S_i S_j)}
  {\sum_{\{S\}} \exp (K\sum_{\langle ij\rangle} \tau_{ij}S_i S_j)}.
      \label{m-def-detail2}
  \end{equation}
As both sides of this equation are independent of the choice of the values
of $\{\sigma_i\}$, we may sum the right hand side over all possible values
of $\{\sigma_i\}$ and divide the result by $2^N$, where $N$ is the total
number of sites, to find
 \begin{equation}
  m= \frac{1}{2^N (2\cosh K_p)^{N_B}}
  \sum_{\{\tau_{ij}\}}\sum_{\{\sigma_i\}} \sigma_i
   \exp (K_p \sum_{\langle ij\rangle} \tau_{ij} \sigma_i \sigma_j)
     \langle S_i \rangle_K.
  \end{equation}
By inserting the identity $1=Z(K_p,\{\tau_{ij}\})/Z(K_p,\{\tau_{ij}\})$
just in front of the sum over $\{\sigma_i\}$
in the summand, we obtain
  \begin{eqnarray}
 \fl
    m= \frac{1}{2^N (2\cosh K_p)^{N_B}}
  \sum_{\{\tau_{ij}\}}\sum_{\{\sigma_i\}}
   \exp (K_p \sum_{\langle ij\rangle} \tau_{ij} \sigma_i \sigma_j)
   \frac{\sum_{\{\sigma_i\}}\sigma_i
    \exp (K_p \sum_{\langle ij\rangle} \tau_{ij} \sigma_i \sigma_j)}
  {\sum_{\{\sigma_i\}}
    \exp (K_p \sum_{\langle ij\rangle} \tau_{ij} \sigma_i \sigma_j)}
     \langle S_i \rangle_K \\
  \lo
  =\frac{1}{2^N (2\cosh K_p)^{N_B}}
  \sum_{\{\tau_{ij}\}}\sum_{\{\sigma_i\}}
   \exp (K_p \sum_{\langle ij\rangle} \tau_{ij} \sigma_i \sigma_j)\,
   \langle \sigma_i \rangle_{K_p}
     \langle S_i \rangle_K 
      \label{2ndlast}\\
   \lo
  =\frac{1}{(2\cosh K_p)^{N_B}}
      \sum_{\{\tau_{ij}\}} \exp (K_p\sum_{\langle ij \rangle}\tau_{ij})
     \langle \sigma_i \rangle_{K_p}
     \langle S_i \rangle_K .
      \label{last}
 \end{eqnarray}
The last line (\ref{last}) can be confirmed by applying the gauge
transformation to (\ref{last}) and using the same argument as above with
gauge invariance
of the product $ \langle \sigma_i \rangle_{K_p} \langle S_i \rangle_K$ in mind
to derive (\ref{2ndlast}).
Equation (\ref{last}) is exactly the definition of the right hand side of
(\ref{magnetization}), which completes the proof.
A similar argument leads to (\ref{C-identity}).


\end{document}